\documentclass[preprint,12pt]{elsarticle}
\usepackage[utf8]{inputenc}
\usepackage{caption}
\usepackage{subcaption}
\usepackage{gensymb}
\usepackage{amsmath}
\usepackage{amsfonts}
\usepackage{amssymb}
\usepackage{float}
\usepackage{lineno}
\graphicspath{{./figures/}}

\begin{document}
\title{The Radioactive Source Calibration System of the PROSPECT Reactor Antineutrino Detector}
\author[wlab]{J.\,Ashenfelter}
\author[wisc]{A.\,B.\,Balantekin}
\author[wlab]{H.\,R.\,Band}
\author[lmc]{C.\,D.\,Bass}
\author[nist]{D.\,E.\,Bergeron}
\author[temple]{D.\,Berish}
\author[llnl]{N.\,S.\,Bowden}
\author[llnl]{J.\,P.\,Brodsky}
\author[hfir]{C.\,D.\,Bryan}
\author[psl]{J.\,J.\,Cherwinka}
\author[llnl]{T.\,Classen}
\author[gt]{A.\,J.\,Conant}
\author[ornl]{D.\,Dean}
\author[hfir]{G.\,Deichert}
\author[bnl]{M.\,V.\,Diwan}
\author[drex]{M.\,J.\,Dolinski}
\author[gt]{A.\,Erickson}
\author[wlab]{B.\,T.\,Foust}
\author[ornl]{M.\, Febbraro}
\author[wlab]{J.\,K.\,Gaison}
\author[ornl,ut]{A.\,Galindo-Uribarri}
\author[ornl,ut]{C.\,E.\,Gilbert}
\author[ornl,ut]{B.\,T.\,Hackett}
\author[bnl]{S.\,Hans}
\author[temple]{A.\,B.\,Hansell}
\author[wlab]{K.\,M.\,Heeger}
\author[drex]{J.\,Insler}
\author[bnl]{D.\,E.\,Jaffe}
\author[temple]{D.\,C.\,Jones}
\author[drex]{O.\,Kyzylova}
\author[drex]{C.\,E.\,Lane}
\author[wlab]{T.\,J.\,Langford}
\author[nist]{J.\,LaRosa}
\author[iit]{B.\,R.\,Littlejohn}
\author[ornl,ut]{X.\,Lu}
\author[iit]{D.\,A.\,Martinez\,Caicedo}
\author[ornl]{J.\,T.\,Matta}
\author[cwm]{R.\,D.\,McKeown}
\author[nist]{M.\,P.\,Mendenhall}
\author[ornl]{P.\,E.\,Mueller}
\author[nist]{H.\,P.\,Mumm}
\author[temple]{J.\,Napolitano}
\author[drex]{R.\,Neilson}
\author[wlab]{J.\,A.\,Nikkel}
\author[wlab]{D.\,Norcini}
\author[nist]{S.\,Nour}
\author[uwloo]{D.\,A.\,Pushin}
\author[bnl]{X.\,Qian}
\author[ornl,ut]{E.\,Romero-Romero}
\author[bnl]{R.\,Rosero}
\author[uwloo]{D.\,Sarenac}
\author[iit]{P.\,T.\,Surukuchi}
\author[wlab]{A.\,B.\,Telles\corref{cor1}}
\ead{arina.telles@yale.edu}
\author[nist]{M.\,A.\,Tyra}
\author[ornl]{R.\,L.\,Varner}
\author[bnl]{B.\,Viren}
\author[iit]{C.\,White}
\author[temple]{J.\,Wilhelmi}
\author[wlab]{T.\,Wise}
\author[bnl]{M.\,Yeh}
\author[drex]{Y.-R.\,Yen}
\author[bnl]{A.\,Zhang}
\author[bnl]{C.\,Zhang}
\author[iit]{X.\,Zhang}

\address[bnl]{Brookhaven National Laboratory, Upton, NY, USA}
\address[drex]{Department of Physics, Drexel University, Philadelphia, PA, USA}
\address[gt]{George W.\,Woodruff School of Mechanical Engineering, Georgia Institute of Technology, Atlanta, GA USA}
\address[iit]{Department of Physics, Illinois Institute of Technology, Chicago, IL, USA}
\address[llnl]{Nuclear and Chemical Sciences Division, Lawrence Livermore National Laboratory, Livermore, CA, USA}
\address[lmc]{Department of Physics, Le Moyne College, Syracuse, NY, USA}
\address[nist]{National Institute of Standards and Technology, Gaithersburg, MD, USA}
\address[hfir]{High Flux Isotope Reactor, Oak Ridge National Laboratory, Oak Ridge, TN, USA}
\address[ornl]{Physics Division, Oak Ridge National Laboratory, Oak Ridge, TN, USA}
\address[temple]{Department of Physics, Temple University, Philadelphia, PA, USA}
\address[ut]{Department of Physics and Astronomy, University of Tennessee, Knoxville, TN, USA}
\address[uwloo]{Institute for Quantum Computing and Department of Physics and Astronomy, University of Waterloo, Waterloo, ON, Canada}
\address[cwm]{Department of Physics, College of William and Mary, Williamsburg, VA, USA}
\address[wisc]{Department of Physics, University of Wisconsin, Madison, Madison, WI, USA}
\address[psl]{Physical Sciences Laboratory, University of Wisconsin, Madison, Madison, WI, USA}
\address[wlab]{Wright Laboratory, Department of Physics, Yale University, New Haven, CT, USA}

 \cortext[cor1]{Correspondence to: Yale Wright Laboratory, 272 Whitney Ave, New Haven, CT 06511, USA.}
\date{\today}

\begin{abstract}
The Precision Reactor Oscillation and Spectrum (PROSPECT) Experiment is a reactor neutrino experiment designed to search for sterile neutrinos with a mass on the order of 1~eV/c$^2$ and to measure the spectrum of electron antineutrinos from a highly-enriched $^{235}$U nuclear reactor.
The PROSPECT detector consists of an 11 by 14 array of optical segments in $^{6}$Li-loaded liquid scintillator at the High Flux Isotope Reactor in Oak Ridge National Laboratory.
Antineutrino events are identified via inverse beta decay and read out by photomultiplier tubes located at the ends of each segment.
The detector response is characterized using a radioactive source calibration system.
This paper describes the design, operation, and performance of the PROSPECT source calibration system.
\end{abstract}


\maketitle


\section{Introduction}

The Precision Reactor Oscillation and Spectrum (PROSPECT) Experiment is a short-baseline, on-surface antineutrino experiment designed to search for sterile neutrinos with a mass on the order of 1~eV/c$^2$ and precisely measure the antineutrino spectrum from $^{235}$U fission daughters~\cite{programpaper}.  Currently operating at the High Flux Isotope Reactor at Oak Ridge National Laboratory, the PROSPECT collaboration is making these measurements to address previously observed reactor neutrino anomalies: the neutrino flux deficit and the deviation from the expected antineutrino reactor spectrum. The first results of PROSPECT are detailed in~\cite{firstresults}.

\begin{figure}[h]
\centering
\includegraphics[width=\textwidth]{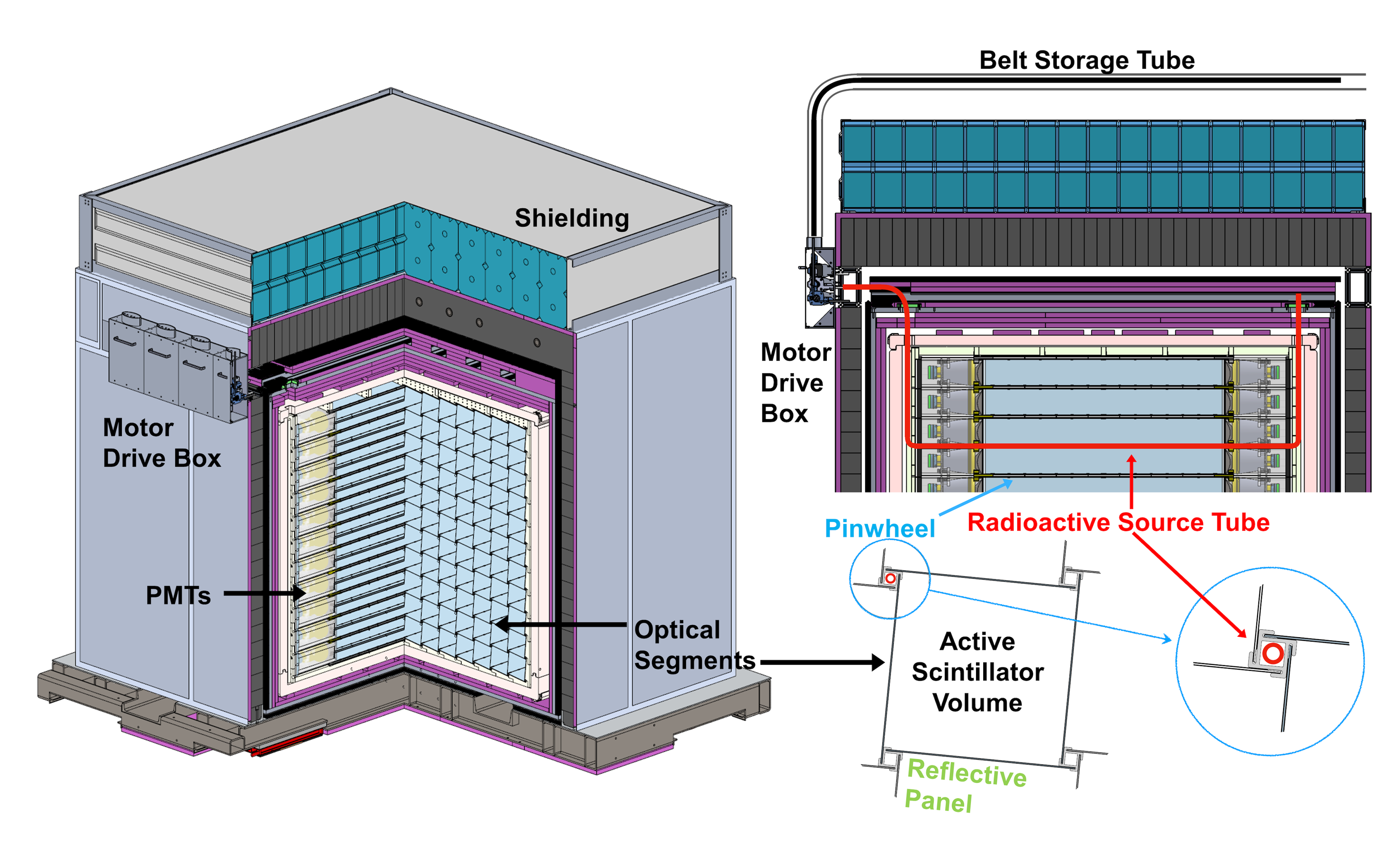}
\caption{(color online) Overview of PROSPECT detector and source calibration system. The cutaway on the left shows detector geometry, and the section view on the top right shows the path of the calibration tube (red) through the detector. The reflective panels (green) that make up the optical segments are clipped together with plastic ``pinwheels," (blue) which are shown on the bottom right and provide a natural location for inserting radioactive sources and other instrumentation.}
\label{fig:overview}
\end{figure}

The PROSPECT detector consists of an optically separated array of 11 by 14 segments, filled with 4 tons of $^{6}$Li-doped liquid scintillator~\cite{scintillator}. A cutaway view of the detector is shown in figure~\ref{fig:overview}. Each optical segment is 117.6~cm long with a cross section of 14.5 by 14.5~cm.  Photomultiplier tubes (PMTs) collect scintillation light at both ends of each segment. Charged particles traveling through the scintillator produce scintillation photons, the number and time structure of which are a function of the energy deposited and the particle type. PROSPECT identifies neutrinos that interact with the scintillator via inverse beta decay (IBD). This occurs when an electron antineutrino from the reactor interacts with a proton, producing a positron and a neutron. The positron ionizes the scintillator and then annihilates with an electron, producing two 511 keV gamma rays. The neutron captures predominantly on $^{6}$Li within approximately 50~$\mu s$, yielding an alpha particle and a triton.  The coincidence of the positron ionization and annihilation with the delayed neutron capture provide the unique signature of an antineutrino.

Accurate reconstruction of these interactions relies on extensive detector calibration, which is achieved by placing sources of radiation in the detector volume. The radioactive source calibration system is used to determine the overall energy scale and nonlinearity, and to study systematic effects such as 511~keV gamma rays escaping from the detector volume and segment-to-segment variations. In addition,  the calibration system is used to determine the detector's neutron response and position reconstruction. This paper describes the source calibration system of the PROSPECT detector; more general details of the detector design and construction can be found in~\cite{nimpaper}.  

The design of the calibration system is naturally coupled to the specific features of the detector.  In order to optically separate the individual segments in PROSPECT, thin reflective panels are clipped together with plastic parts (``pinwheels'') to form a square lattice with a $5.5\degree$ tilt~\cite{opticalgrid}.  This results in small inactive volumes at the corners of the individual segments.  Figure~\ref{fig:overview} includes a cross section of a PROSPECT segment, showing the reflective panels and the plastic pinwheels.  The pinwheels have a square inner hollow that is approximately 10~mm across, ideal for inserting small radioactive sources, optical fibers, and thermometers through tubes to a variety of points inside the detector.  Calibration systems that use tubes to deploy radioactive sources are not unusual in particle physics detector designs (\cite{ZeplinCalib}, \cite{MajoranaCalib}, \cite{CuoreCalib}), but this system introduces a few unique features that are described in the sections below.

\section{System Components}

The source calibration system uses motor drive units to deploy small radioactive sources along the lengths of the optical segments.   There are a few key requirements that motivated the system's design: 

\begin{itemize}
\item
Chemical compatibility with the liquid scintillator
\item
Minimized dead mass inside the detector
\item
Full detector coverage
\item
Compact deployment system to fit reactor site constraints
\item
Remote operation and monitoring of source positions
\item
Production within a short timescale, compatible with detector commissioning 
\end{itemize}

To achieve these goals, a mechanism was developed that deploys radioactive sources through low-friction tubes using a timing belt driven by stepper motors.  This system is single-ended, and each calibration tube can be operated independently. The motor drive box (left of figure~\ref{fig:overview}) contains the motors that drive the timing belts in and out of the detector, and the end of each belt is mounted with a small radioactive source capsule. Custom pulleys, casings, and connectors were developed to interface between the tubes, belts, and motors.  3D printing and other rapid fabrication techniques were employed to speed prototyping and production while reducing costs.  

\begin{figure}[h]
\centering
\begin{minipage}{.62\linewidth}
\centering
\includegraphics[width=.99\textwidth]{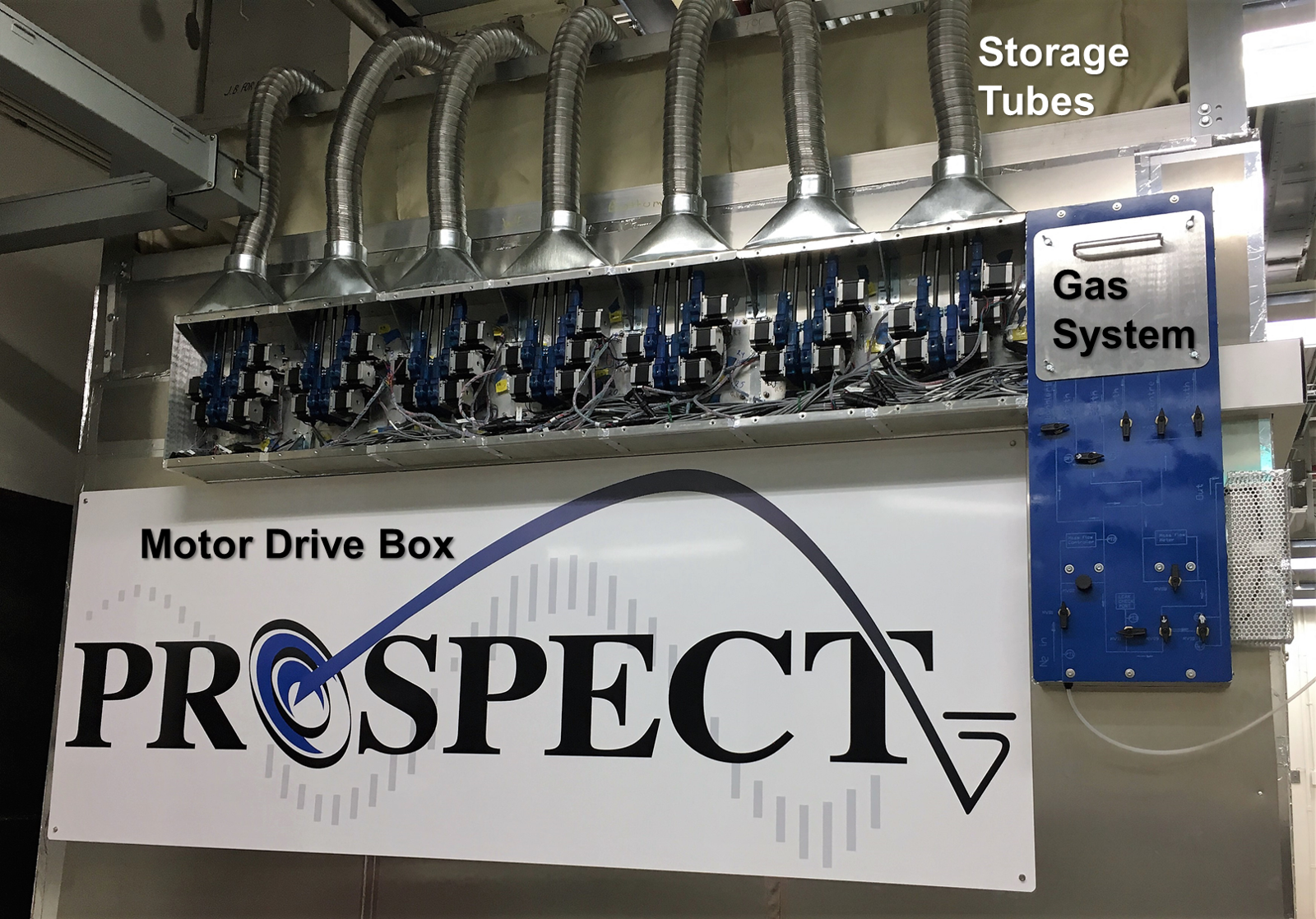}
\end{minipage}
\begin{minipage}{.371\linewidth}
\centering
\includegraphics[width=.99\textwidth]{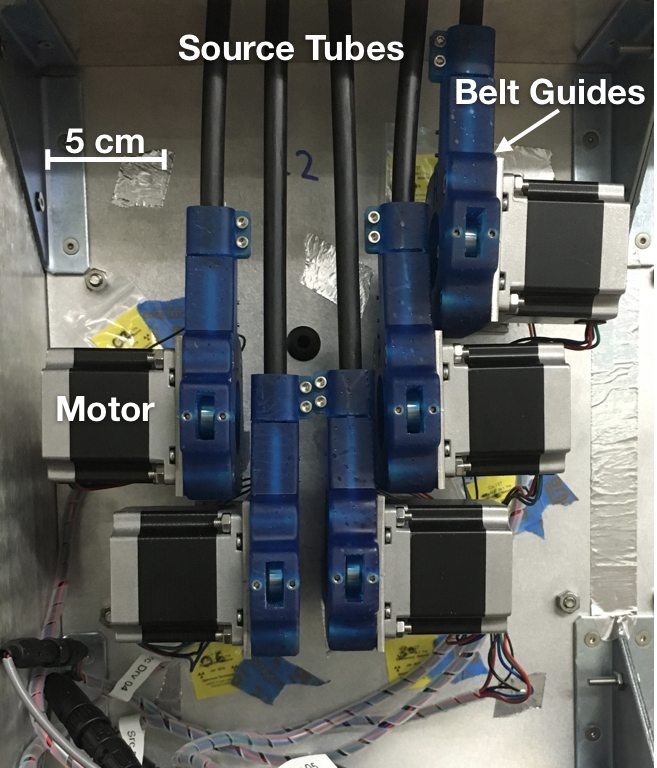}
\end{minipage}
\caption{Left: The fully installed source calibration system.  The 35 motor drives fit into a box approximately 2~m wide, and extend from the face of the detector by only 14~cm.  The storage tubes are routed through the aluminum ducts over the top of the detector.  During operation, the box is covered with panels to ensure that light does not travel into the detector volume.  The panel to the right of the motors is the cover gas system for the scintillator. Right: One set of motor drives, servicing one column of the detector. }
\label{fig:calibphoto}
\end{figure}

During detector calibrations, one or more of the motors drive a source out to a given position along the length of the tube. The position is specified and monitored through a slow control web interface, and the motors are controlled with Arduino\footnote{Electronics prototyping and development platform. Arduino AG, Turin, Italy. \\ Note: Certain trade names and company products are mentioned in the text or identified in illustrations in order to adequately specify the experimental procedure and equipment used. In no case does such identification imply recommendation or endorsement by the National Institute of Standards and Technology, nor does it imply that the products are necessarily the best available for the purpose.} micro-controllers. When calibrations are complete, the sources are retracted all the way to the motor drive box, where they may be exchanged or stored for regular data-taking. When the sources are retracted, the excess belt routes into the storage tubes above the detector. There is sufficient shielding between the radioactive sources and the active part of the detector that no significant backgrounds are introduced when the sources are in their retracted ``home" position. Photos of the system are shown in figure~\ref{fig:calibphoto} and the components are described in the following sections.

\subsection{Source Calibration Tubes}
As the PROSPECT detector array is fully immersed in liquid scintillator, all the materials used must be chemically compatible.  This was addressed by threading PTFE\footnote{Polytetrafluoroethylene.} tubes through the centers of the pinwheels to provide a compatible barrier between the inserted items and the scintillator. Both ends of the tubes are located outside of the scintillator volume to provide complete isolation.  Furthermore, since the corner spaces between the segments are already a necessary part of the assembly, no additional inactive volumes are introduced.
 
There are thirty-five tubes dedicated to source deployment arranged into a 5 by 7 grid.  The tubes have a 3/8" outer diameter and 1/4" inner diameter.  They are arranged such that a given optical segment is both adjacent to a source tube and two segments away from a different source tube.  This placement is valuable for understanding uniformity of response across segments as well as any large scale variation.  

The tubing was provided by the manufacturer in rolls, so it was annealed in sections in a custom oven to straighten it.   After a final cut during detector assembly, each tube was installed and then measured using a calibrated and labeled length of timing belt. Prior bench tests were done to determine if bends in the tubing would affect the measurement, and no discernible difference between straight or bent tubes was found.

The far ends of each quintet of tubes in each column are connected via a manifold to a single tube that routes back to the motor drive box. If a source capsule were to unexpectedly detach from the belt during deployment, blowing compressed air through this tube pushes the source back out to the motor drive box, where it can be easily retrieved.

\subsection{Sources and Source Capsules}
A comprehensive calibration requires a selection of both gamma and neutron sources.  PROSPECT currently deploys $^{137}$Cs, $^{60}$Co, $^{22}$Na, $^{252}$Cf, and AmBe. These provide calibrations for a critical range of energies over which the scintillator is non-linear, as well as for positron annihilation and neutrons (see table~\ref{tab:sources}).

\begin{table}[H]
\caption{Calibration sources and their uses.}
\centering
\begin{tabular}{|c|c|c|c|c|}
\hline 
\textbf{Source} & \textbf{Type} & \textbf{$\gamma$ Energy (MeV)} & \textbf{Primary Purpose} & \textbf{Rate} \\
\hline 
$^{137}$Cs & gamma  & 0.662 & segment comparison & 0.1 $\mu$C \\  
\hline
$^{22}$Na & gamma & 2x 0.511, 1.275 & positron, edge effects & 0.1 $\mu$C \\ 
\hline
$^{60}$Co & gamma  & 1.173, 1.332 & energy scale& 0.1 $\mu$C \\ 
\hline 
$^{252}$Cf & neutron & 2.223 (n-H capture) & neutron response & 866 n/s \\ 
\hline 
AmBe & neutron & -- & neutron response & 70 n/s\\
\hline 
\end{tabular} 
\label{tab:sources}
\end{table}

\begin{figure}[H]
\centering
\includegraphics[width=0.75\textwidth]{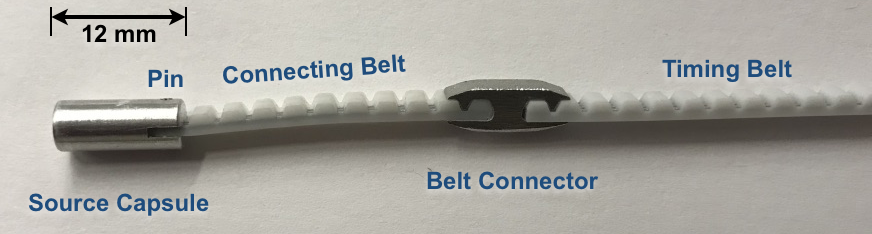}
\caption{Source capsule attached by belt connector to the timing belt. The cylindrical capsule is manufactured with a slot to insert the belt. A steel spring pin is pressed through small holes in this slot, capturing the belt by piercing through middle of it. The pin is compressed in this procedure, attaching very securely to the belt. The belt slides sideways in and out of the tightly fitted belt connector.  All the source capsules are identical, so each are individually laser etched to identify them.}
\label{fig:capsule}
\end{figure}

The sources are encapsulated into small aluminum cylinders, each sealed with a set-screw and epoxy (see figure~\ref{fig:capsule}). Each capsule is etched with a unique identification number that is recorded in the slow control database.  The capsule dimensions are the largest that can safely move through a 1.5" radius $90\degree$ bend of PTFE tubing, which is the required curvature for tubes exiting the ends of the optical segments. The capsule attaches to the timing belt (section~\ref{sec:belt}) with a stainless steel spring pin.  This mounts the source permanently to a short section of timing belt, ensuring that it stays firmly attached to the belt while it is deployed. This short section is joined to the longer stretch of belt via custom-made belt connectors. The longer stretch of belt remains installed in its own tube, while the shorter section with the capsule may be switched by sliding the belt out of the connector. This scheme facilitates swapping sources between tubes because only a small section of belt has to be moved between the motor drives. In addition, it minimizes the risk of damaging the source capsule (as well as radiation exposure) because the capsule itself is never manipulated after it is initially mounted on the belt.

\subsection{Timing Belt}
\label{sec:belt}

The timing belt used to push the sources through the detector is one of the more unusual features of this system.  This specific timing belt was chosen because it experiences little friction inside the tube and places the source capsules to reproducible positions, regardless of tube length or curvature.  A timing belt, as opposed to a rod or string, also moves the sources in discrete steps, which simplifies keeping track of source positions.  It was found experimentally that as long as the combination of belt and tubing is correct, pushing the belt into the tube works very well. The belt must be the right width and stiffness to avoid buckling in the tube.  This timing belt is a 3~mm wide, AT3 pitch (distance from one tooth to the next is 3~mm), polyurethane belt reinforced with steel cords\footnote{BRECOflex Co., LLC, Eatontown, New Jersey, U.S.A. }. 

\subsection{Belt Guide and Pulley}

The timing belt is driven by a 3D printed pulley on a NEMA 23 stepper motor\footnote{NEMA 23 Stepper Motor. 5.4v 1.5A 1.16Nm. Stepper Online, OMC Corp. Lim., Nanjing City, China. }.  The pulley has 33 teeth, and slides onto the D-shaped motor shaft. An adjustable spring-loaded jockey wheel presses the belt against the pulley to keep the belt from skipping teeth during normal operation.  The spring tension is adjusted so that the belt {\it can} skip if there is an unusual amount of force applied to the belt, such as in the case of a kink or obstruction in the source or storage tube.  

\begin{figure}[h]
\centering
\begin{minipage}{0.43\linewidth}
\includegraphics[width=0.95\textwidth]{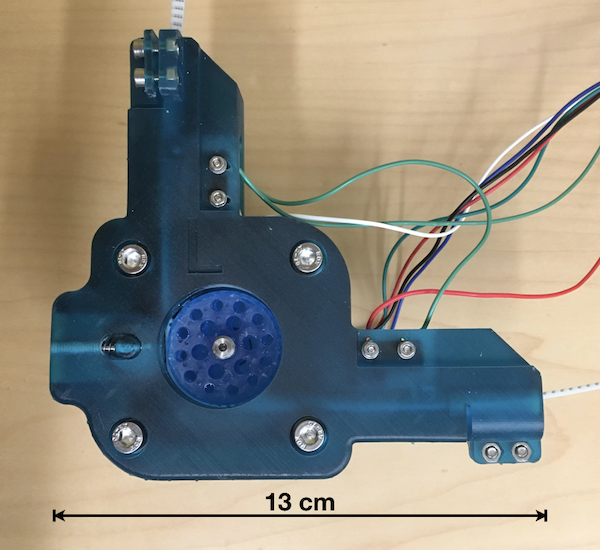}
\end{minipage}
\begin{minipage}{0.55\linewidth}
\includegraphics[width=0.99\textwidth]{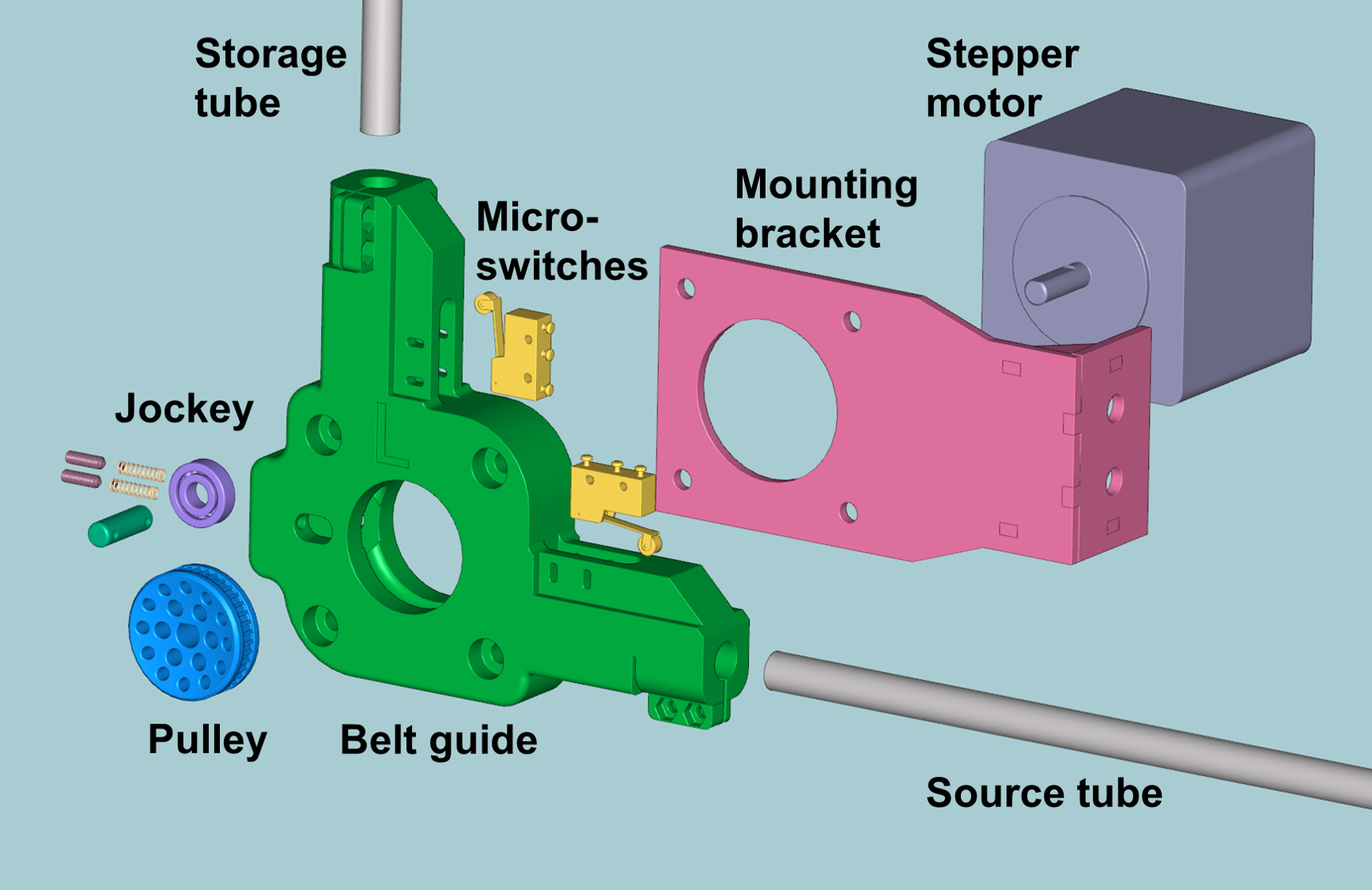}
\end{minipage}
\caption{Left: 3D printed belt guide and pulley. Right: Exploded CAD view of belt guide and pulley assembly.}
\label{fig:beltguide_exploded}
\end{figure}

A 3D printed belt guide houses the drive assembly and guides the belt from the source tube to the pulley, and out to the storage tube.  A photo and an exploded view of the entire unit are shown in figure~\ref{fig:beltguide_exploded}.  The belt guide contains one micro-switch that stops the motor if the source capsule approaches the pulley while retracting, acting both as a safety feature and as the home position of the source capsule.  The belt guide is designed such that the tube sits in a well-defined location relative to this home position.  An additional micro-switch on the storage tube side prevents the belt from extending beyond the pulley.   The distance between the belt and the switch contacts can be adjusted such that normal vibration of the belt does not trip the switch, but the capsule or connector always does.  The switches are electrically connected and the micro-controllers programmed such that if any connection is broken, or if a cable is unplugged, the motor will not run.
	
	The timing belt pulleys and belt guides were designed specifically for this system and 3D printed using a UV-cured resin\footnote{Formlabs Tough V4. Formlabs, Somerville, Massachusetts, U.S.A.}.  The overall design and shape of the belt guides was driven by space requirements at the reactor site.  As the motor drive box is located outside the detector shielding, it is the most prominent part of the the detector on that side, so keeping the assembly as compact as possible was a key constraint. Figure~\ref{fig:overview} illustrates where the features and components of the calibration system are located in and on the detector.

\subsection{Electronics and Control System}
\label{sec:control}

\begin{figure}[H]
\centering
\includegraphics[width=\linewidth]{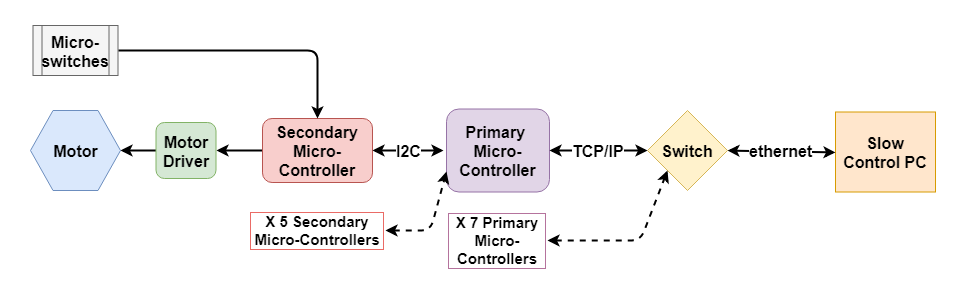}
\caption{Diagram of the control system for source calibration.}
\label{fig:control_diagram}
\end{figure}

The drive units are operated through the PROSPECT slow control interface by an array of inexpensive micro-controllers\footnote{Arduino Controllers. DEV-11166, DEV-11061, DEV-11113. Sparkfun Electronics, Boulder, Colorado, U.S.A.}. A diagram of the control scheme is shown in figure~\ref{fig:control_diagram}. Each motor and corresponding set of micro-switches are controlled by one ``Arduino Pro Mini," which is labeled as the secondary micro-controller. These are connected in groups of five, and each group is controlled by a single primary micro-controller, an ``Arduino Ethernet," via the I$^2$C protocol. Thus, seven primary controllers coordinate the entire system, each communicating with the slow control computer via TCP/IP over ethernet cables. This strategy for distributing the computing load simplifies  the software and ensures that the locations of the belts are always known. 

Positioning is accomplished by having each secondary controller continuously count the number of pulses that are sent to the motor drivers, as well as the direction that the motor rotates with each pulse.  Each pulse corresponds to 123.75~$\mu$m of belt movement, so the micro-controller can report the location of the source at any time. When the belt is fully retracted, the belt connector trips the micro-switch that keeps it from over-retracting, defining the capsule's home position.  The absolute position of the radioactive source is determined by counting the number of steps that the belt has extended and comparing that to physical measurements of the tube, which were measured upon assembly.  

\section{Performance and Detector Response}

Before installing the calibration system in PROSPECT, a series of long term feasibility tests were conducted. Five full scale calibration tubes were arranged as they would be in the detector but mounted to a frame such that the system could be observed in operation.  All five of these units were run for 12,000 cycles over 3 weeks.  The source tubes are semi-transparent, so the location of the test capsules could be observed directly.  The location of the capsule returned to within 2~mm on subsequent extensions, and absolute positioning along the source tube was estimated to be within 2~cm based on measuring the extended source tube with a measuring tape. In the actual detector the source cannot be externally observed, but since the test setup is identical to a full column the detector's geometry, these values are adequate for describing the full system's positioning precision.

The PROSPECT calibration system is currently operational in the PROSPECT detector and calibration data is collected regularly.  The overall response of the detector is studied in three main ways: energy reconstruction, position reconstruction, and spatial dependence. The energy scale, neutron efficiency, and gamma position reconstruction studies described below are essential as inputs for the overall PROSPECT physics analysis. Together, all these calibration analyses take full advantage of the system's features and provide valuable cross-checks for the detector simulation.

\subsection{Energy Reconstruction}
Events of a specific energy are produced by deploying gamma sources into the detector as described above. Gamma events are identified by pulse-shape discrimination (PSD) and their energy is reconstructed by the amount of scintillation detected.  PMT signals are converted to an energy scale by calculating pulse integral spectra and comparing with deposited energy spectra generated by a Monte Carlo (MC) simulation. 

\begin{figure}[H]
\centering
\includegraphics[width=\linewidth]{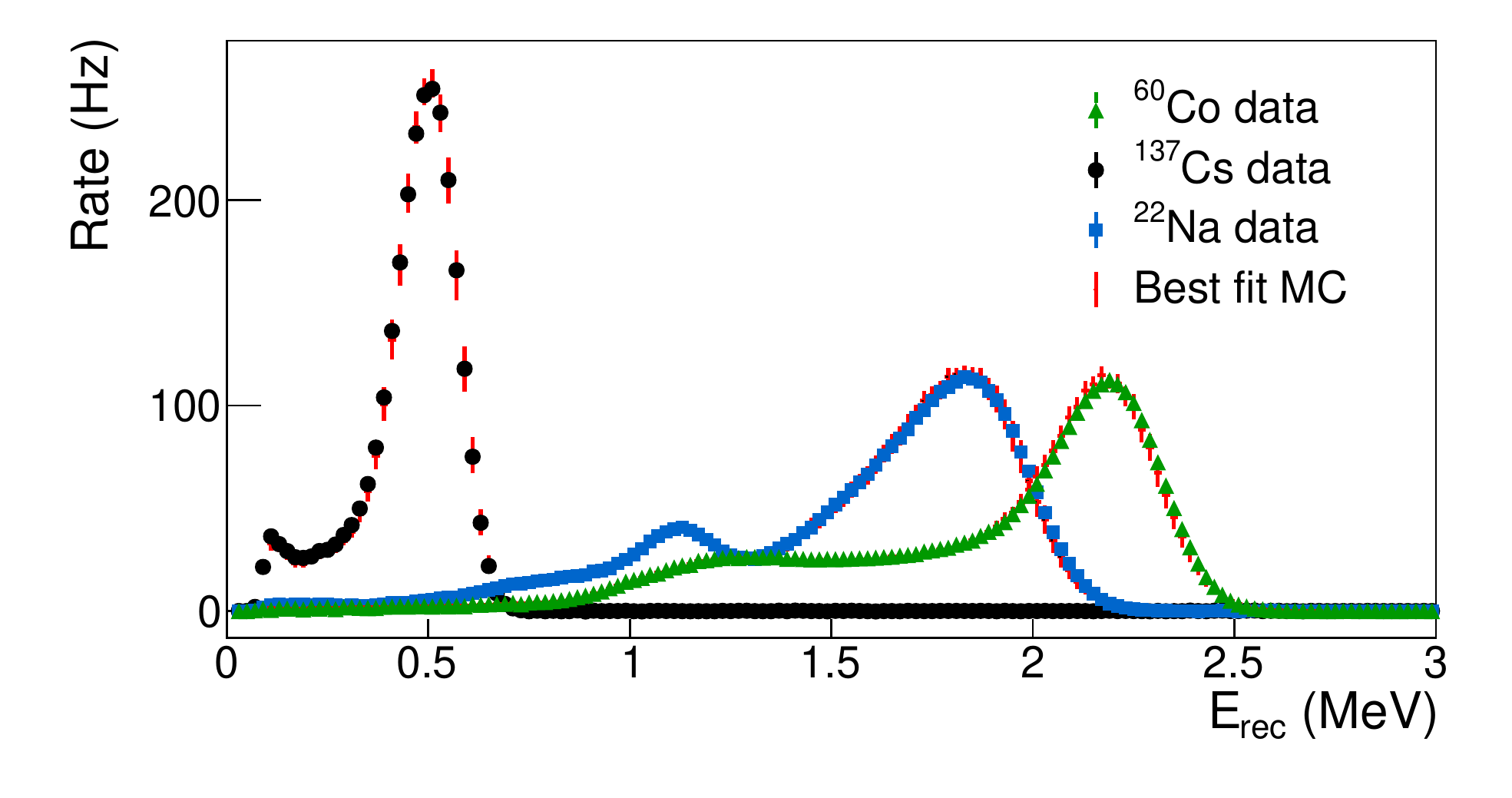}
\caption{Energy spectra of $^{60}$Co, $^{137}$Cs, and $^{22}$Na obtained using data from the entire detector, showing excellent agreement with the simulation. Note that $^{60}$Co is analyzed as the sum of the two gamma rays that it emits.  Overall, the reconstructed visible energies are shifted from their true energies due to multiple scatters, non-linear light production of the scintillator, and energy loss in dead material. The true energy of peaks are 0.66 MeV  for $^{137}$Cs, 1.27 MeV for the $^{22}$Na single gamma, 2.30 MeV for the sum of the three $^{22}$Na gammas, and 2.50 MeV for the sum of the two  $^{60}$Co gammas.}
\label{fig:escale}
\end{figure}

The energy spectra of the entire suite of gamma sources is shown in figure~\ref{fig:escale}, demonstrating gamma energy coverage up to 2.5~MeV.  Data-driven models, which include energy non-linearity and scale, are used to generate MC spectra used for these studies. A simultaneous fit between data and MC of the peaks shown in figure 6, the $\beta$~spectrum from $^{12}$B and the 2.2~MeV gamma peak from the neutron capture on H is performed and minimized. The best fit model produces a relative data-to-MC difference of $<1\%$.  This calibration is vital in characterizing the energy scale of the detector, constraining the nonlinearity model, and calculating an accurate detector response model for both the sterile neutrino search and $^{235}$U spectrum analysis. Further information on the fitting procedure and the use of this analysis in PROSPECT results can be found in~\citep{firstresults}. 

\subsection{Position Reconstruction along Segment Lengths}

The position of any given particle interaction within an optical segment of the detector is determined using the timing difference and the charge ratio between pulses of PMTs opposite each other. During source calibrations, there is additional position information provided by the control system described above. Comparing the deployed position to its reconstruction helps to characterize the detector's position response, especially as a function of segment length and location. In particular, the distance between the prompt and delayed IBD signal is a useful metric for identifying neutrino signals and determining fiducial volume. In addition, having multiple sources enables studies of position resolution across a range of energies.

\begin{figure}[H]
\centering
\includegraphics[width=\textwidth]{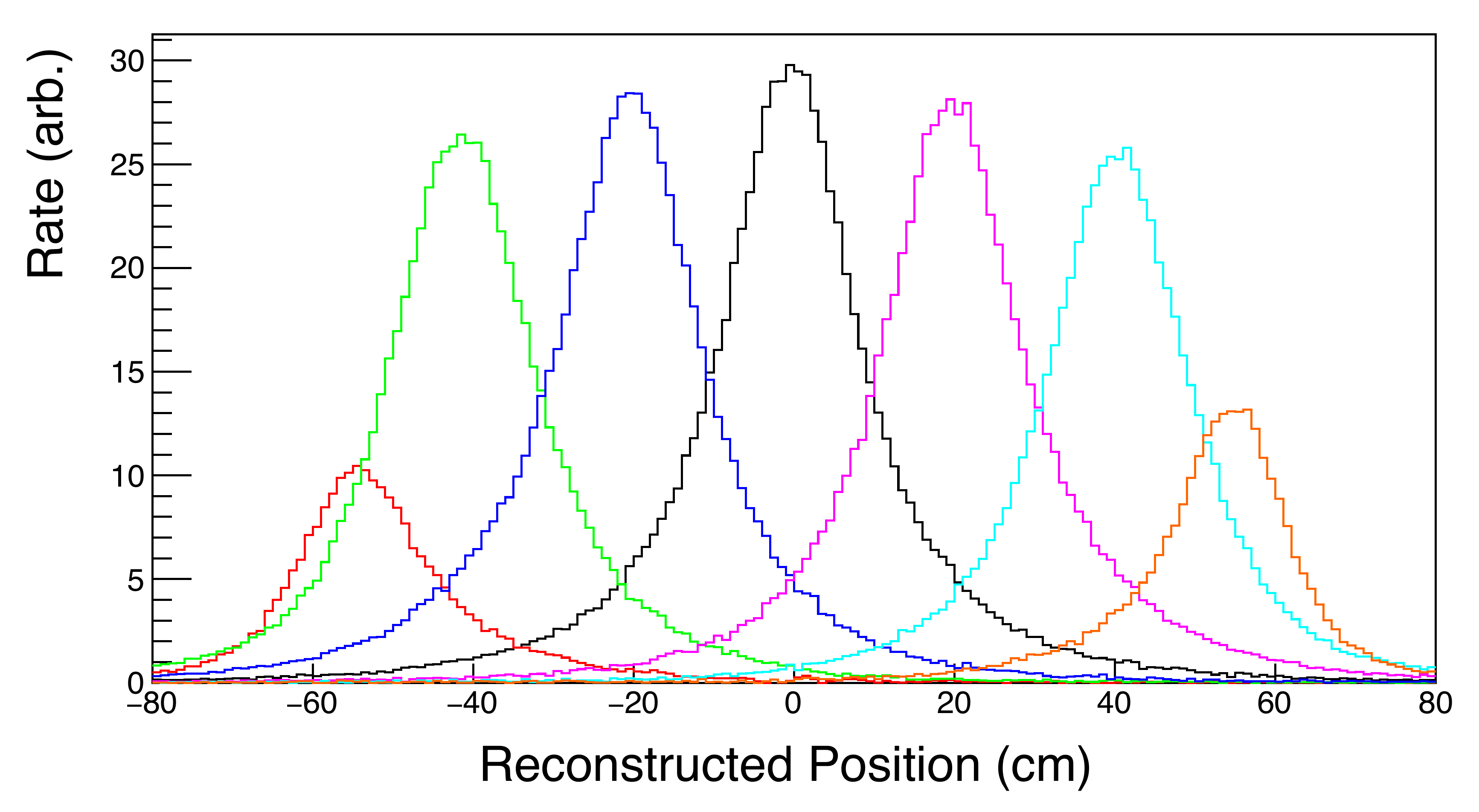}
\caption{Histogram of $^{137}$Cs deployments along the length of an optical segment. Note that the optical segment is 117.6~cm long, so calibrations taken at $\pm$60~cm are at the very end of the segments. This accounts for the 60~cm peaks being relatively low, since the source radiates mostly outside the detection volume. It is worth noting that these edge distributions are sensitive enough to the ends of the active volume that shifting the source until the peaks are the same height could possibly be used as a means of adjusting for any uncertainties in absolute positioning incurred when tube lengths were measured upon assembly. }
\label{fig:zscan_plot}
\end{figure}

Figure~\ref{fig:zscan_plot} is an example of a typical calibration which consists of deploying the source at 20~cm increments along the segment and analyzing the distributions of counts along its length. Each peak signifies an individual deployment of $^{137}$Cs to a location along the segment. The width of the peaks is the combined effect of the natural spread of the radiation from the source, as well as the detector's position resolution. $^{137}$Cs is used for this study because of its simple decay, but deploying other sources enables analyses of position resolution over a range of energies.

\subsection{Spatial Dependence of Detector Response}

The response of the PROSPECT detector to particle interactions varies across different segments, and along the lengths of each segment. This is partly due to the fact that in the center of the detector, gammas or other particles are typically completely captured, whereas near the edges of the detector, they can escape the active volume. Calibrations are performed to quantify the effects of the detector geometry on the data. In addition, calibrations allow any other slight variations across segments to be accounted for.

\begin{figure}[H]
\centering
\includegraphics[width=\linewidth]{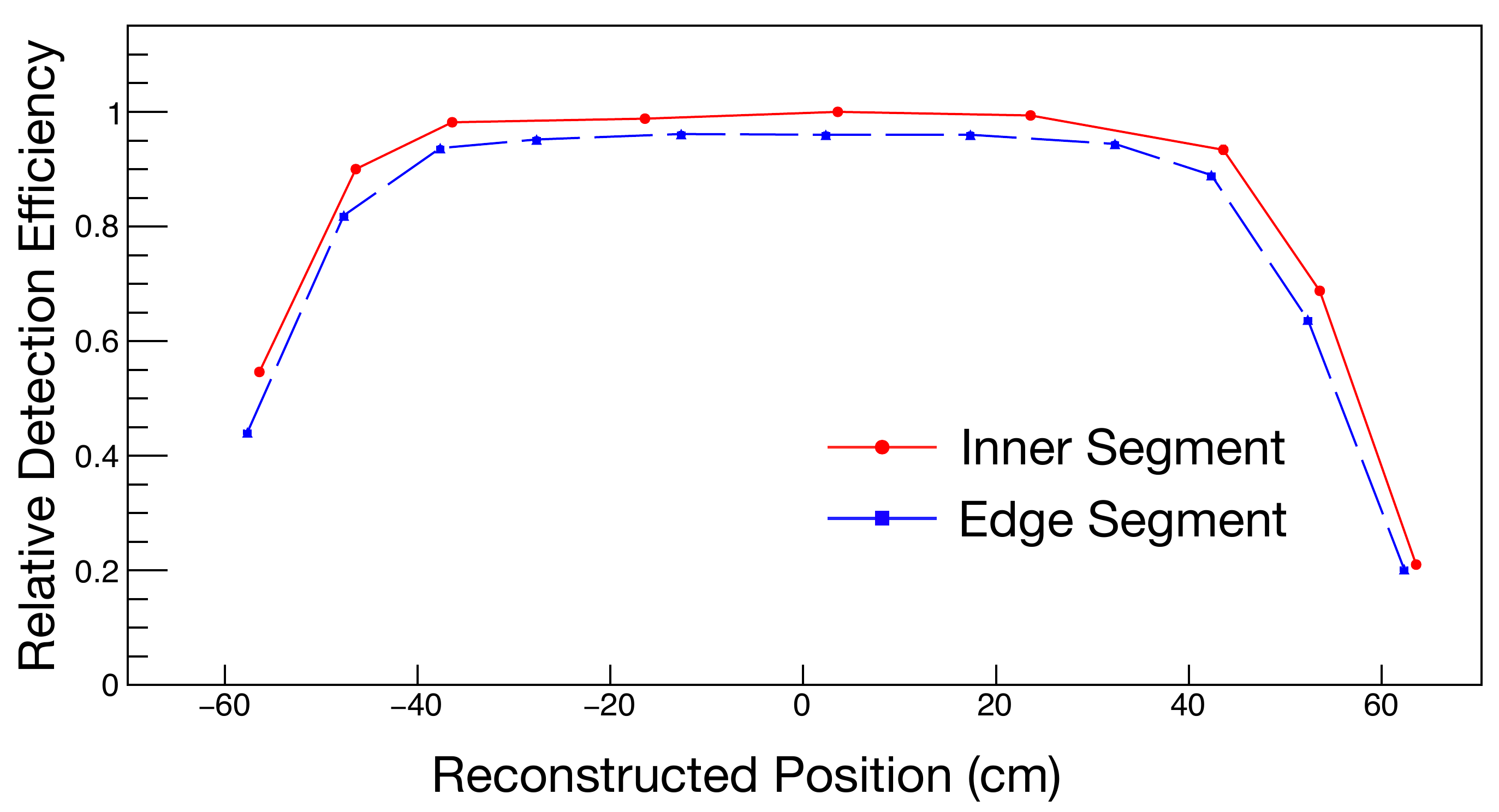}
\caption{Neutron detection efficiency for deployments of $^{252}$Cf in two different source tubes. Each data point represents the efficiency measured as the source was moved to that location along the length of a source tube. The inner segment line is higher than the edge segment because fewer neutrons escape the detector when the source is near the center.}
\label{fig:neutrons}
\end{figure}

Neutron detection efficiency is an important metric to map across the detector, since neutron capture on $^{6}$Li is used to identify inverse beta decay. $^{252}$Cf is deployed at various locations to characterize neutron response, an example of which is shown in figure~\ref{fig:neutrons}.

The neutron detection efficiency varies across the optical array as neutrons leave the active volume near the detector edges. This accounts for the difference in the two tubes shown in figure~\ref{fig:neutrons}: higher efficiency is from a source placed near the center, while lower efficiency is found closer to the edge. Likewise, the efficiency falls off steeply at the ends of the segments (when the source is near $\pm$~60 cm) because many of the neutrons exit the volume before they can be captured by the scintillator. The behavior of inner segments vs. edge segments is important as an input to the simulation, in turn informing the sterile neutrino oscillation analysis.

\begin{figure}[H]
\centering
\includegraphics[width=\linewidth]{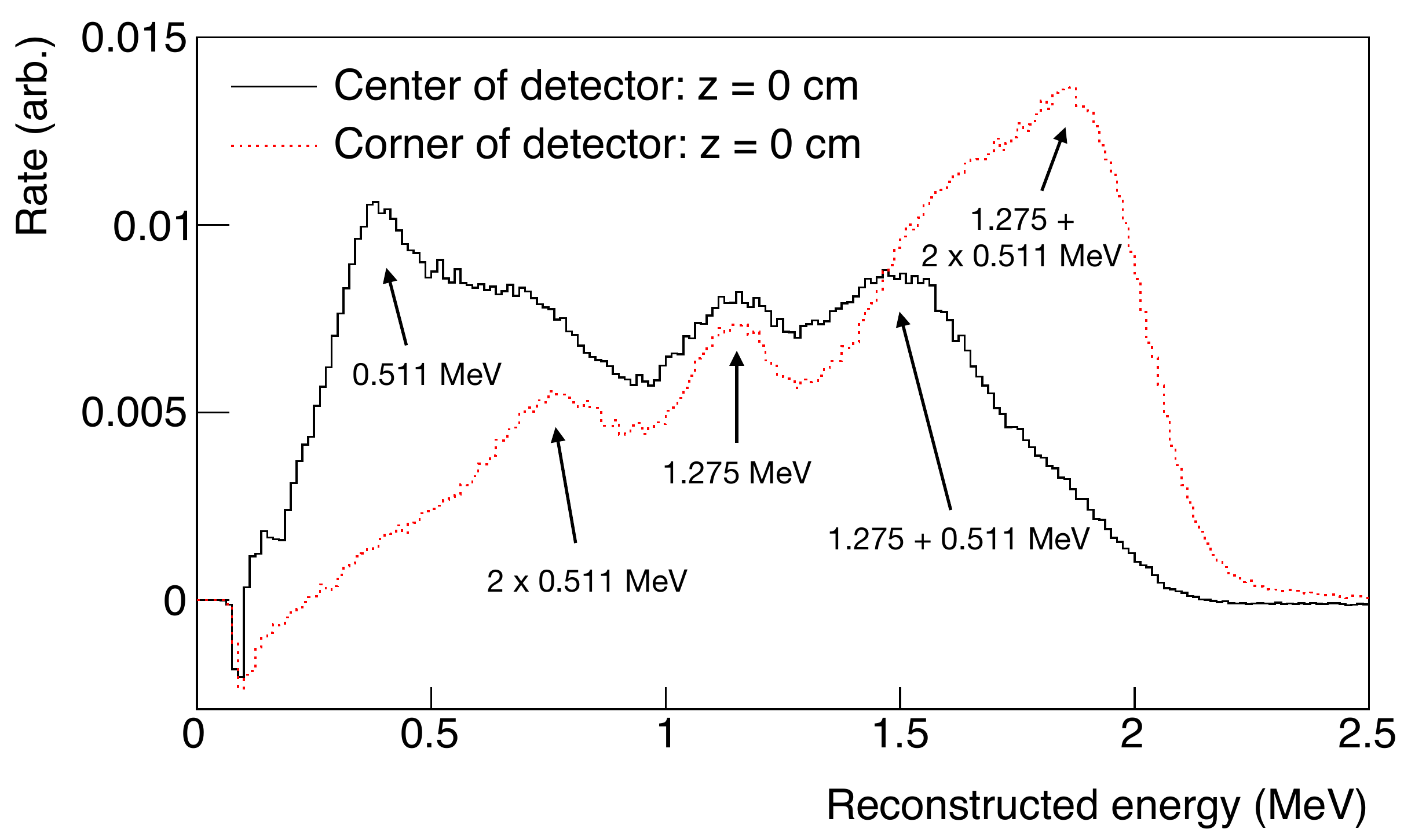}
\caption{Illustration of the difference between a $^{22}$Na spectrum measured for a deployment at the edge vs. the center of the detector.  The peaks are labeled with their true energies. The reconstructed energy scale of the entire detector is shifted from the true energy due to multiple scatters and non-linear light production of the scintillator.}
\label{fig:edge_eff}
\end{figure}

Similarly, source calibration is used to study the behavior of the two 0.511~MeV gamma rays emitted from positron annihilation near detector edges. The measured difference for deployments of a positron source near the center and near the edges is shown in figure~\ref{fig:edge_eff}.

The $^{22}$Na source is used to study edge effects for gamma rays because of its convenient decay scheme: it emits a 1.275~MeV gamma and a positron which captures in the capsule, releasing two 0.511~MeV gamma rays with opposite momenta. This allows for various combinations of gamma rays to be observed at the center and edge of the detector. For example, the sum of all three emitted gammas is prominent in the center spectrum because their interaction lengths are spanned by a few optical segments. In the edge spectrum, this peak is absent due to one of the 0.511 MeV rays escaping the volume. These studies benchmark the systematic uncertainty from missing energy affecting energy reconstruction differently across the detector.

\begin{figure}[H]
\centering
\includegraphics[width=\linewidth]{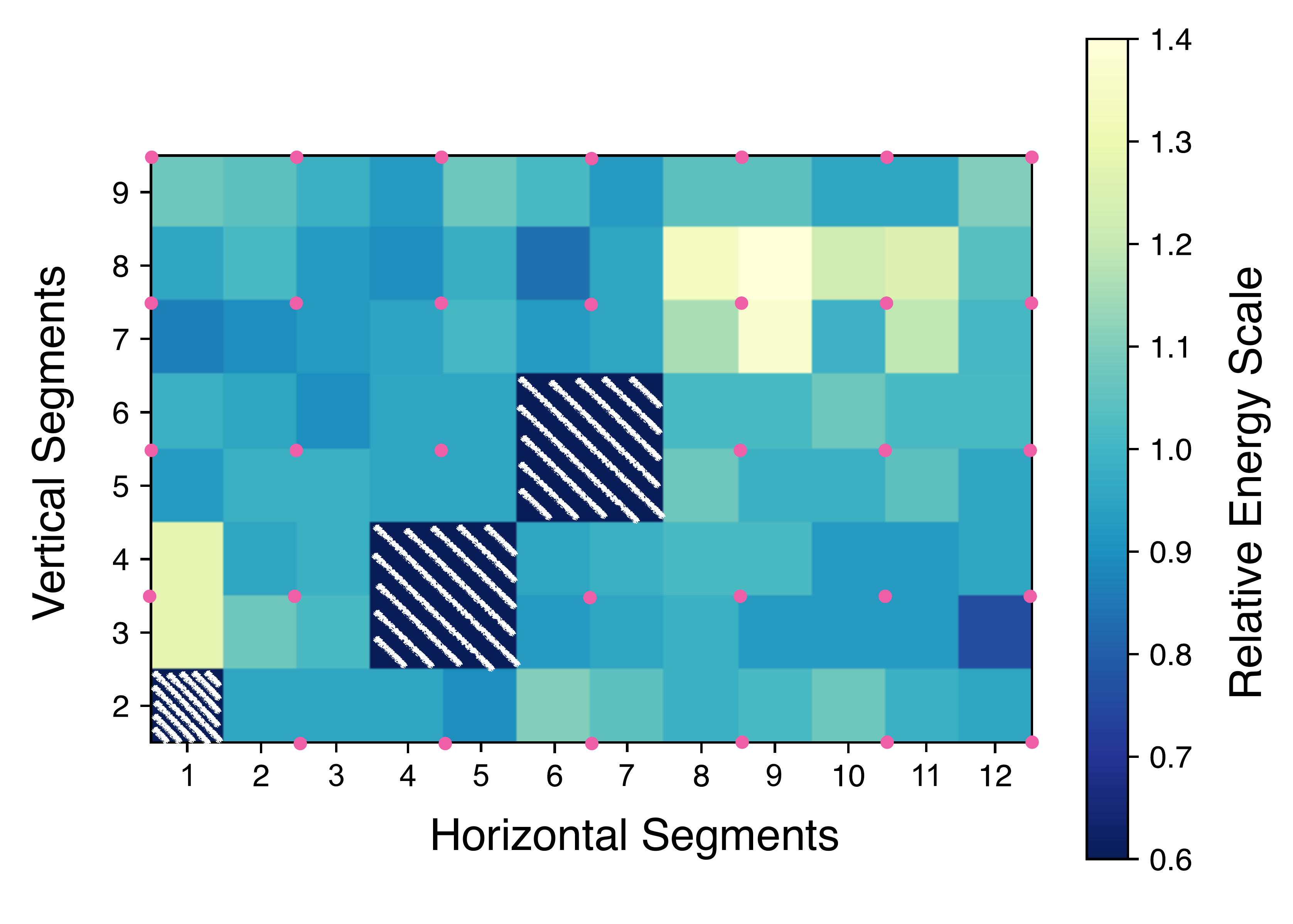}
\caption{(color online) The uncalibrated segment-to-segment variation in energy across the central region of the detector. The differences are great because this occurs before any calibrations are applied in the analysis to account for the segment  and PMT differences. The sources used were $^{60}$Co and $^{137}$Cs.  The dots indicate the locations of the deployed sources.  The hatched areas signify that no source was deployed in those segments due to a problem inserting it.  See main text for details.}
\label{fig:segtoseg}
\end{figure}

Finally, overall segment-to-segment energy scale variation can be studied across the detector by deploying the same type of gamma source in different source tubes, using the fact that a single source illuminates the four segments that surround it. Figure~\ref{fig:segtoseg} shows the results from multiple source deployments throughout the entire optical array. The energy response of each segment was fitted with the same simulated template of the spectrum to quantify variation. These variations can then be used to examine if there are any systematic issues with calibration that are spatially dependent. This analysis is sensitive to gain differences, offsets in absolute source positioning, and any other slight differences between the segments.  The three hatched areas in figure~\ref{fig:segtoseg} surround three source deployment tubes where it was not possible to insert a source.  The most likely reason for this is that these three tubes were kinked or crushed during the final assembly of the detector.

\section{Conclusion}

The PROSPECT source deployment system was designed to introduce gamma and neutron sources directly into all regions of the detector's active volume.  It takes advantage of the details of the detector's segmented design and has been used for multiple successful detector calibrations.   The system has several original features, such as the use of timing belts, which make it possible to deploy a small radioactive source to precise and repeatable locations along the optical segments.  The control system uses multiple inexpensive micro-controllers working in harmony to drive and continuously monitor the status of the sources.  Rapid prototyping techniques were used extensively to speed development and enabled the production of a compact system. This specific design makes it easy to map the response to various energies and particles across the entire detector; a vital requirement to reach the physics goals of the PROSPECT experiment.  Periodic calibration runs will continue to track detector performance over time.

\section{Acknowledgements}

This material is based upon work supported by the following sources: US Department of Energy (DOE) Office of Science, Office of High Energy Physics under Award No. DE-SC0016357 and DE-SC0017660 to Yale University, under Award No. DE-SC0017815 to Drexel University, under Award No. DE-SC0008347 to Illinois Institute of Technology, under Award No. DE-SC0016060 to Temple University, under Contract No. DE-SC0012704 to Brookhaven National Laboratory, and under Work Proposal Number  SCW1504 to Lawrence Livermore National Laboratory. 
This work was performed under the auspices of the U.S. Department of Energy by Lawrence Livermore National Laboratory under Contract DE-AC52-07NA27344 and by Oak Ridge National Laboratory under Contract DE-AC05-00OR22725. 
Additional funding for the experiment was provided by the Heising-Simons Foundation under Award No. \#2016-117 to Yale University. 

J.G. is supported through the NSF Graduate Research Fellowship Program and A.C. performed work under appointment to the Nuclear Nonproliferation International Safeguards Fellowship Program sponsored by the National Nuclear Security Administration’s Office of International Nuclear Safeguards (NA-241). 
This work was also supported by the Canada  First  Research  Excellence  Fund  (CFREF), and the Natural Sciences and Engineering Research Council of Canada (NSERC) Discovery  program under grant \#RGPIN-418579, and Province of Ontario.

We further acknowledge support from Yale University, the Illinois Institute of Technology, Temple University, Brookhaven National Laboratory, the Lawrence Livermore National Laboratory LDRD program, the National Institute of Standards and Technology, and Oak Ridge National Laboratory.
We gratefully acknowledge the support and hospitality of the High Flux Isotope Reactor and Oak Ridge National Laboratory, managed by UT-Battelle for the U.S. Department of Energy.

\bibliographystyle{./style/h-physrev}
\bibliography{ref}

\end{document}